# Pseudo Dirac Nodal Sphere: Unusual Electronic Structure and Material Realization

Jianfeng Wang[1], Yizhou Liu[2], Kyung-Hwan Jin[3], Xuelei Sui[2,1], Wenhui Duan[2,4], Feng Liu[3,4], and Bing Huang[1]

[1] *Beijing Computational Science Research Center, Beijing 100193, China*
[2] *Department of Physics and State Key Laboratory of Low-Dimensional Quantum Physics, Tsinghua University, Beijing 100084, China*
[3] *Department of Materials Science and Engineering, University of Utah, Salt Lake City, Utah 84112, USA*
[4] *Collaborative Innovation Center of Quantum Matter, Beijing 100084, China*

**Abstract**

**Topological semimetals (TSMs) in which conduction and valence bands cross at zero-dimensional (0D) Dirac nodal points (DNPs) or 1D Dirac nodal lines (DNLs), in 3D momentum space, have recently drawn much attention due to their exotic electronic properties. Here we generalize the TSM state further to a higher-symmetry and higher-dimensional pseudo Dirac nodal sphere (PDNS) state, with the band crossings forming a 2D closed sphere at the Fermi level. The PDNS state is characterized with a spherical backbone consisting of multiple crossing DNLs while band degeneracy in between the DNLs is approximately maintained by weak interactions. It exhibits some unique electronic properties and low-energy excitations, such as collective plasmons different from DNPs and DNLs. Based on crystalline symmetries, we theoretically demonstrate two possible types of PDNS states, and identify all the possible band crossings with pairs of 1D irreducible representations to form the PDNS states in 32 point groups. Importantly, we discover that strained $M$H$_3$ ($M$= Y, Ho, Tb, Nd) and Si$_3$N$_2$ are materials candidates to realize these two types of PDNS states, respectively. As a high-symmetry-required state, the PDNS semimetal can be regarded as the "parent phase" for other topological gapped and gapless states.**

The rise of topological insulator [1,2] has brought the field of topological state to the center stage of condensed matter physics. Recent attentions have been focused on topological semimetals (TSMs), which can support quasiparticles either analogous to elementary particles in high-energy physics or unknown before [3-7]. To date, the well-known TSMs include Dirac, Weyl, and nodal-line semimetals [5-14]. The Dirac semimetals [5-7] have zero-dimensional (0D) band crossings, i.e., the Dirac nodal points (DNPs), whose Fermi surface consists of isolated points in the Brillouin zone (BZ) [upper panel, Fig. 1(a)]. The low-energy excitations (LEEs) of DNP semimetals have some unique properties such as chiral anomaly and surface states with Fermi arcs. The nodal-line semimetals [6-14] feature 1D band crossings at Fermi surface with closed Dirac nodal lines (DNLs) in the BZ [upper panel, Fig. 1(b)]. The DNL semimetals host special drumhead surface states, which provide an important platform to realize strong electron correlation effect. Very recently, a nodal surface is also proposed, with the band crossing points forming a 2D plane [15,16].

Unlike DNPs and DNLs, conceptually it is also possible that the linear band crossing occurs on a 2D closed surface [17,18], forming a Dirac nodal sphere (DNS) or pseudo DNS (PDNS, characterized with a spherical backbone made of multiple crossing DNLs and approximate band degeneracy in between the DNLs, see details below) at Fermi energy, as shown in the upper panel of Fig. 1(c). On a DNS/PDNS, each point is a crossing point between two bands with linear dispersion along the surface normal direction, which can be expressed as

$$H(k') = \hbar v_F k' \sigma_z, \qquad (1)$$

where $k' = k - k_0$ is the component of wave vector normal to the Fermi surface, $k_0$ is the radius of DNS, $v_F$ is the Fermi velocity, and $\sigma_z$ is Pauli matrice denoting the two crossing bands.

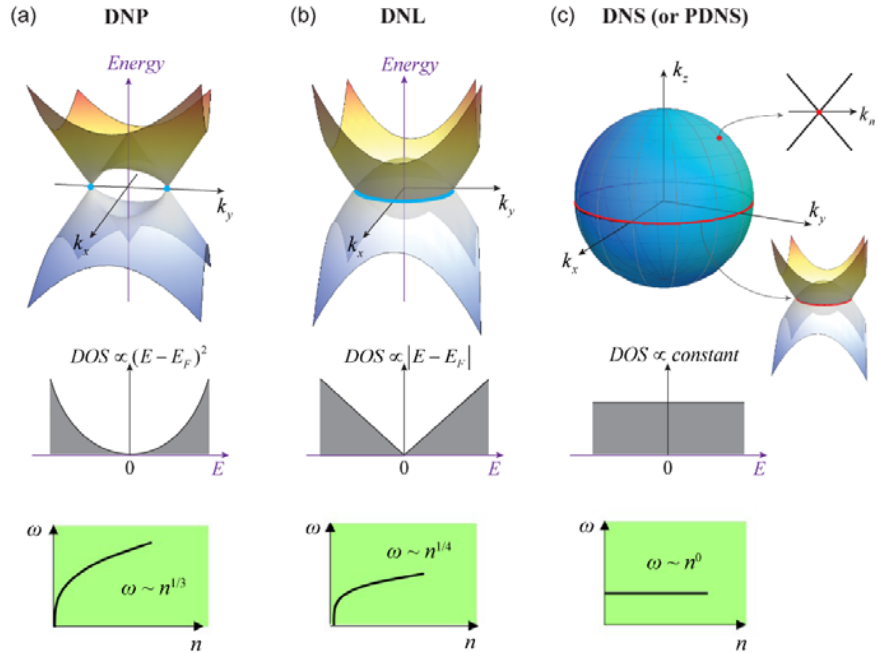

Figure 1. Comparison between (a) Dirac nodal point (DNP), (b) Dirac nodal line (DNL) and (c) Dirac nodal sphere (DNS) or pseudo DNS (PDNS) for their distinct Fermi surface geometries (upper panels), density of states (middle panels) and carrier density dependent plasmon frequency (bottom panels) around the Fermi level.

Since the LEE dimensionality of DNS is fundamentally different from that of DNP (DNL), the DNS semimetal can possess very unique electronic properties. For example, it has a significantly different density of states (DOS): $DOS \propto (E - E_F)^2$ for a DNP [middle panel, Fig. 1(a)], $DOS \propto |E - E_F|$ for a DNL [middle panel, Fig. 1(b)], and $DOS \propto constant$ [19] for a DNS [middle panel, Fig. 1(c)]. Another important feature of DNS is its unusual collective modes of plasma oscillation. It is known that the long-wavelength ($q \to 0$) plasmon frequency in DNP semimetals is distinct from the classical plasmons in parabolic metals in all dimensions [20]. Here, we find the long-wavelength plasmon frequency of DNS to be significantly different from DNP and DNL semimetals, even though all of them belong to 3D Dirac systems [19-21]:

$$\hbar\omega_{\text{DNP}} = \sqrt{\frac{2ge^2\hbar v_F}{3\pi\kappa}}(6\pi^2 n/g)^{1/3} + O(q^2), \tag{2a}$$

$$\hbar\omega_{\text{DNL}} = \sqrt{\frac{2\pi e^2 \hbar v_F}{\kappa}(1 + \sin^2\theta)}(\pi g k_0 n)^{1/4} + O(q^2), \tag{2b}$$

$$\hbar\omega_{\text{DNS}} = \sqrt{\frac{4ge^2\hbar v_F k_0^2}{3\pi\kappa}} + O(q^2), \tag{2c}$$

where $n$ is the carrier density, with the degeneracy factor $g$ and dielectric constant $\kappa$. The long-wavelength plasmon frequency for all 3D Dirac systems is independent of the wave vector $q$ in the leading term of Eq. (2), similar to normal plasmons in 3D metals with a collective excitation gap. However, the plasmon frequency of DNS is also independent of the carrier density ($\omega_{\text{DNS}} \propto n^0$) [bottom panel, Fig. 1(c)], which is distinctly different from that of DNP ($\omega_{\text{DNP}} \propto n^{1/3}$) [bottom panel, Fig. 1(a)], DNL ($\omega_{\text{DNL}} \propto n^{1/4}$) [bottom panel, Fig. 1(b)] and parabolic metals ($\omega_{\text{PM}} \propto n^{1/2}$). The constancy of $\omega_{\text{DNS}}$ arises from the 1D Dirac-like DNS carrier with a "relativistic" effective mass, $m_c = E_F/v_F^2$, which depends on carrier density $n$ to cancel out the corresponding $n^{1/2}$ term in the parabolic metals [$\hbar\omega_{\text{PM}} = (4\pi e^2 n/\kappa m)^{1/2}$] [19]. Therefore, the DNS fermion exhibiting unique electronic properties and field responses can be recognized as a new type of fermion beyond the DNP/DNL paradigm.

One intriguing question is how to realize this novel DNS state in realistic materials. Although the (Weyl) nodal sphere state has been theoretically proposed based on global symmetries [17,18], it is too difficult, if not impossible, to be realized in real crystals having discrete point group symmetries. Here, we present an effective approach to generate PDNS state in accessible crystalline symmetries. In general, a band crossing located on high-symmetry lines/planes is stable against band repulsion only when the wavefunctions belong to different eigenstates of some crystalline symmetry operation. For an ideal DNS, the band degeneracy should occur at an arbitrary momentum point (say $P$ point) on the sphere. But generally the coupling between two crossing bands at $P$ cannot be strictly avoided. Interestingly, near some high-symmetry $k$ points, we discover that under some appropriate conditions the special crystalline symmetries will only allow for the high-order interaction terms (HITs) of $k$ between two crossing bands, which can be sufficiently weak and hence negligible. In this case, a PDNS state forms, whose LEEs are same as an ideal one, albeit only a key subset of crossing points (DNLs) formed as the spherical backbone of the PDNS are topologically protected.

We identify two sets of crystalline symmetries under time-reversal symmetry (TRS) to realize the desired PDNS states: type-I for inversion plus two mirror ($\hat{P}+2\hat{M}$) symmetries and type-II for three mirror ($3\hat{M}$) symmetries. Importantly, we identify all the possible band crossings with pairs of 1D irreducible representations (IRRs) to form these two types of PDNS states in 32 point groups. Employing first-principles calculations, we further show that $M$H$_3$ ($M$= Y, Ho, Tb, Nd) and Si$_3$N$_2$ are type-I and type-II PDNS semimetals under certain strains, respectively. They both have drumhead surface states independent of surface orientations and constant long-wavelength plasmon frequency independent of $q$ and $n$.

We start with a general two-band model in a system with TRS, written as [11-14]:
$$H(\mathbf{k}) = g_0(\mathbf{k})\sigma_0 + g_1(\mathbf{k})\sigma_x + g_2(\mathbf{k})\sigma_y + g_3(\mathbf{k})\sigma_z, \tag{3}$$
where $\sigma_0$ is the identity matrix, and $\sigma_{x,y,z}$ are Pauli matrices for the two bands of interest near the Fermi level. Here we ignore the spin degree of freedom. The TRS requres $g_{0,1,3}(\mathbf{k})$ [$g_2(\mathbf{k})$] to be real even [odd] functions of $\mathbf{k}$. The eigenvalues of Eq. (3) are $E(\mathbf{k}) = g_0(\mathbf{k}) \pm \sqrt{g_1^2(\mathbf{k}) + g_2^2(\mathbf{k}) + g_3^2(\mathbf{k})}$. Since we are interested in the band crossing, the $g_0(\mathbf{k})$ term, which represents an overall kinetic energy, can be neglected henceforth. The degeneracy of eigenvalues (or band crossing) requires $g_1(\mathbf{k}) = g_2(\mathbf{k}) = g_3(\mathbf{k}) = 0$. The vanishing $g_{1,2,3}(\mathbf{k})$ at discrete points, a continuous line or a sphere in the BZ will create DNP, DNL or DNS (PDNS), respectively. Importantly, the crystalline symmetries pose additional constraints on $g_i(\mathbf{k})$, which further enforces band degeneracy. Overall, we discover that there are at least two different types of symmetry requirements to realize the PDNS states.

The type-I PDNS has $\hat{P}+2\hat{M}$ symmetries. If there are two bands having opposite parities at a high-symmetry point, e.g., Γ point, $\hat{P}$ can be chosen as $\hat{P} = \sigma_z$. Combining the constraints on $g_i(\mathbf{k})$ from $\hat{P}$ symmetry and TRS, one can obtain that $g_1(\mathbf{k}) = 0$ and $g_2(\mathbf{k})$ [$g_3(\mathbf{k})$] is an odd [even] function of $\mathbf{k}$, which can lead to a band crossing on a DNL [11-14]. We further consider two additional (different) mirror symmetries, without loss of generality, $\hat{M}_x$ and $\hat{M}_y$. If these two bands also have opposite mirror eigenvalues for both $\hat{M}_x$ and $\hat{M}_y$, $\hat{M}$ can be chosen as $\hat{M} = \sigma_z$. The two additional $\hat{M}$ symmetries further require $H(\mathbf{k})$ to satisfy: $k_x \leftrightarrow -k_x$; $k_y \leftrightarrow -k_y$. Consequently, the linear term in $g_2(\mathbf{k})$ has to vanish, so that $g_2(\mathbf{k})$ will have only the third or higher-order terms of $k$, e.g., $g_2(\mathbf{k}) = \delta k_x k_y k_z$ to the third order. Up to the lowest order of $\mathbf{k}$, $g_3(\mathbf{k})$ can be written as $g_3(\mathbf{k}) = M - B\mathbf{k}^2$, considering the isotropy around the high-symmetry point, where $\mathbf{k}^2 = k_x^2 + k_y^2 + k_z^2$. Then, the Hamiltonian under $\hat{P}+2\hat{M}$ symmetries can be written as:
$$H(\mathbf{k}) = (M - B\mathbf{k}^2)\sigma_z + \delta k_x k_y k_z \sigma_y. \tag{4}$$
The condition for $g_3(\mathbf{k}) = 0$ can be satisfied only if $MB > 0$, which is nothing but the band inversion condition. Then, $\hat{P}+2\hat{M}$ will strictly create three crossing nodal lines in $k_{x,y,z} = 0$ planes [11,12]. Away from the three planes, there would be a gap induced by HIT of $g_2(\mathbf{k})$, but it can be sufficiently tiny and negligible for small $k$ near the high-symmetry point. Consequently, the band crossings can extend to form a PDNS. Around a crossing point, the LEE quasiparticles can be described by Eq. (1) with $k_0 = \sqrt{M/B}$ and $v_F = -2\sqrt{MB}$.

The minimum symmetries required for the type-I PDNS are $\hat{P}+2\hat{M}$ plus TRS. Meanwhile, the two inverted bands at the high-symmetry point should belong to two different 1D IRRs, i.e., *R$_1$* and *R$_2$*, which have the opposite parities and mirror eigenvalues. It is emphasized that the required symmetric mirror planes for PDNS can be more than 2, e.g., 3, 4 or even 6. Applying theses criteria to 32 point groups, we identify that 6 point groups can host type-I PDNS, and all the assoiated possible pairs of 1D IRRs are listed in Table I.

The type-II PDNS has $3\widehat{M}$ symmetries. One may take three mirrors as $3\hat{\sigma}_v$ of $C_{3v}$ point group, which are related with each other by $C_3$ rotational symmetry. If the two crossing bands have opposite eigenvalues for $3\hat{\sigma}_v$, $\widehat{M}$ can be written as $\hat{\sigma}_v = \sigma_z$. It is noted that the three mirror reflections will transform $\boldsymbol{k}$ but keep $k_x^2 + k_y^2$ and $k_z$ unchanged. Then, up to the third order of $\boldsymbol{k}$, the Hamiltonian of Eq. (3) will, similar to type-I PDNS, retain only the $g_3(\boldsymbol{k})\sigma_z$ and $g_2(\boldsymbol{k})\sigma_y$ terms:

$$H(\boldsymbol{k}) = (M - B\boldsymbol{k}^2)\sigma_z + \delta(k_x^3 - 3k_x k_y^2)\sigma_y. \tag{5}$$

Once again, strictly speaking, it creates three crossing DNLs, which are related with each other by $C_3$ rotational symmetry. However, away from the three planes, the small gap induced by HIT of $g_2(\boldsymbol{k})$ can be neglected near the high-symmetry point. Thus, we obtain the PDNS under the type-II symmetry constraints.

The minimum symmetries required for the type-II PDNS are $3\widehat{M}$ plus TRS. At the high-symmetry point, the two inverted bands with two different 1D IRRs should have opposite mirror eigenvalues. Also, the required symmetric mirror planes for the type-II PDNS can be more than 3, e.g., 4 or 6. Applying this criterion to 32 point groups, we determine that 9 point groups can potentially host type-II PDNS, and all the associated pairs of 1D IRRs are listed in Table I.

Table I. Two different types of PDNS realized by different point groups with all the possible 1D IRRs, and the proposed materials (without SOC effect), where $M$H$_3$ ($M$= Y, Ho, Tb, Nd), Tl$_5$Se$_2$Br, Tl$_4$PbTe$_3$, Tl$_4$SnTe$_3$ and Si$_3$N$_2$ require strains to realize PDNS states.

| PDNS type | Point group | 1D IRRs of two bands $\{R_1, R_2\}$ | Materials |
|---|---|---|---|
| Type-I | $D_{2h}$ | $\{A_g, A_u\}$, $\{B_{ig}, B_{iu}\}$; $i$=1,2,3 | |
| | $D_{4h}$ | $\{A_{ig}, A_{iu}\}$, $\{B_{ig}, B_{iu}\}$, $\{A_{ig(u)}, B_{ju(g)}\}$; $i,j$=1,2 | LaN, CaTe |
| | $D_{3d}$ | $\{A_{ig}, A_{iu}\}$; $i$=1,2 | YH$_3$, HoH$_3$, TbH$_3$, NdH$_3$ |
| | $D_{6h}$ | $\{A_{ig}, A_{iu}\}$, $\{B_{ig}, B_{iu}\}$, $\{A_{ig(u)}, B_{ju(g)}\}$; $i,j$=1,2 | |
| | $T_h$ | $\{A_g, A_u\}$; | |
| | $O_h$ | $\{A_{ig}, A_{ju}\}$; $i,j$=1,2 | |
| Type-II | $C_{4v}$ | $\{A_1, A_2\}$, $\{B_1, B_2\}$; | |
| | $D_{4h}$ | $\{A_{1g(u)}, A_{2g(u)}\}$, $\{B_{1g(u)}, B_{2g(u)}\}$; | Tl$_5$Se$_2$Br, Tl$_4$PbTe$_3$, Tl$_4$SnTe$_3$ |
| | $C_{3v}$ | $\{A_1, A_2\}$; | |
| | $D_{3d}$ | $\{A_{1g(u)}, A_{2g(u)}\}$; | |
| | $C_{6v}$ | $\{A_1, A_2\}$, $\{B_1, B_2\}$, $\{A_i, B_j\}$; $i,j$=1,2 | |
| | $D_{3h}$ | $\{A_1', A_2'\}$, $\{A_1'', A_2''\}$, $\{A_1', A_1''\}$, $\{A_2', A_2''\}$; | |
| | $D_{6h}$ | $\{A_{1g(u)}, A_{2g(u)}\}$, $\{B_{1g(u)}, B_{2g(u)}\}$, $\{A_{ig(u)}, B_{jg(u)}\}$; $i,j$=1,2 | |
| | $T_d$ | $\{A_1, A_2\}$; | β-Si$_3$N$_2$ |

|  | $O_h$ | $\{A_{1g(u)}, A_{2g(u)}\}$ | α-Si$_3$N$_2$ |

Next, we discuss the topological properties of PDNS. For an ideal DNS semimetal, its topological invariant can be defined on a 0D point enclosing manifold [16-18,22]. Considering two momentum points $k_{in}$ and $k_{out}$ located anywhere inside and outside the ideal DNS, its topological invariant can be defined as $\Delta c = [c(k_{in}) - c(k_{out})]/2$, where $c(k) = \sum_{n\in\text{occ}}\langle u_n(k)|\hat{X}|u_n(k)\rangle$ is a quantum number of symmetry operator $\hat{X}$ for all the occupied bands. However, our PDNS is not an ideal one so that $c$ cannot be well defined at arbitrary $k$ point; instead it needs to be defined within a plane that contains the loop formed by DNLs. Since the DNLs are underlied by the crystal symmetries as we discussed above, one can selectively choose those high-symmetry $k$ points accordingly. For type-I PDNS that has inversion symmetry, $c$ can be defined as the sum of parity for every occupied band at the time-reversal invariant point ($\hat{X} = \hat{P}$); for type-II PDNS, $c$ can be defined as the sum of mirror eigenvalues at the mirror-invariant plane ($\hat{X} = \hat{M}$).

We emphasize that the nontrivial (nonzero) $\Delta c$ defined here cannot protect the whole PDNS against being gapped under a symmetry preserving perturbation, but it can protect the existence of multiple crossing NLs (a necessary condition for achieving the PDNS state). Furthermore, if a perturbation preserves all the required symmetries and maintains the weak band inversion, the band degeneracy of the whole PDNS will be kept. In addition, the zero codimensionality of PDNS cannot, in principle, guarantee any boundary state [16]. The surface states if generated will interact with the bulk states not to be localized on the boundary. However, the drumhead surface states arising from the multiple crossing DNLs are protected on the boundary, except they may appear somewhat fuzzy due to overlapping with the bulk states.

It is noted that we did not include the spin degree of freedom in our PDNS model discussions. All the predicted materials listed in Table I are strictly crossing-nodal-line semimetals with an extremely tiny energy gap (< 2 meV) at a general band crossing point $P$ (induced by HITs of $k$) [19,23]. Without SOC effect, all the candidates listed in Table I can be treated as PDNS semimetals in terms of their LEE properties. For LaN, CaTe, Tl$_5$Se$_2$Br, Tl$_4$PbTe$_3$, and Tl$_4$SnTe$_3$ (Table I), however, the SOC effects are sufficiently strong to reduce the PDNS phase to DNP (or topological insulator) phase [19]. Interestingly, for $M$H$_3$ ($M$= Y, Ho, Tb, Nd) and α-/β-Si$_3$N$_2$, their PDNS phases (under certain strains) are robust against the SOC effect [19], as demonstrated in the following discussions.

Metal hydrides have been studied extensively for superconductivity and metal-insulator transition under pressure [24-26]. YH$_3$ adopts the HoD$_3$ structure [27] having the space group $P\bar{3}c1$ (No. 165), as shown in Fig. 2(a). It has inversion symmetry, threefold rotation symmetry, and three glide planes related by $C_3$ rotation. It is a normal semiconductor whose conduction band minimum (CBM) and valence band maximum (VBM) at the Γ point belong to the $A_{2g}$ and $A_{2u}$ representations of $D_{3d}$ (without SOC effect) [19], respectively. Based on our PDNS model (Table I), it is expected to have a type-I PDNS when its $A_{2g}$ and $A_{2u}$ bands are weakly crossed. Indeed, we found that YH$_3$ can be transformed into a gapless PDNS semimetal when a compressive uniaxial strain ($\varepsilon_c < -3.8\%$) is applied along $c$ axis [19], as shown in Fig. 2(b). As shown in Fig. 2(c), the gapless band crossing maintains along any arbitrary $k$ direction around Γ, with a negligible gap (< 0.5 meV) induced by HITs. We note that the HITs between the two crossing bands are related with the band inversion strength. The condition for the HITs to be negligibly small can always be guaranteed by an appropriate $\varepsilon_c$ [19]. Given the opposite parities of $\hat{P}$ and $\hat{M}$ eigenvalues of three glide planes [labelled in Fig. 2(c)], the band crossing for type-I PDNS is approximately protected by $D_{3d}$ symmetry with a calculated $\Delta c$=1. The calculated constant DOS in the energy range of nearly linear dispersion [Fig. 2(c)] agrees well with that in Fig. 1(c). The spherical Fermi surface of strained YH$_3$ formed by the band crossing is shown in Fig. 2(d), and its size $k_0$ can be tuned by $\varepsilon_c$ [19]. Since the band crossing is not exactly located at the Fermi energy, the Fermi surface presents hole (electron) pockets near (away from) Γ$MK$ plane. These results are not affected by the SOC effect [19], as reflected by a tiny SOC gap (< 1.5 meV).

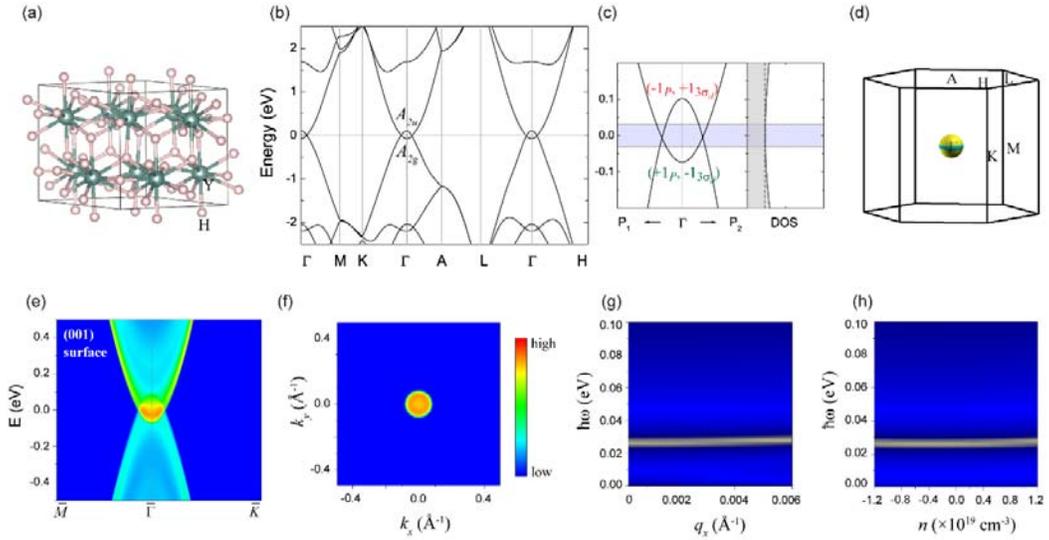

Figure 2. (a) Crystal structure of YH$_3$. (b) Band structure of YH$_3$ under $\varepsilon_c = -6\%$ (without SOC) using HSE06 calculations. (c) Left panel: magnified band structure along two arbitrary directions around Γ ($P_1$ and $P_2$ are two arbitrary $k$ points in BZ), where the opposite eigenvalues of parity and glide planes for two crossing bands are labelled. Right panel: DOS, where the dashed line denotes a constant DOS. (d) The Fermi surface of YH$_3$ in BZ. Cyan (yellow) surface denotes hole (electron) pockets at the Fermi level. (e) and (f) Surface projected bands and Fermi surfaces for (001) surface of YH$_3$. (g) and (h) Plasmon frequency as a function of wave vector $q_x$ and carrier density $n$, respectively.

Drumhead surface states arising from the crossing nodal lines are shown in Figs. 2(e) and (f). Due to the interaction with the projected bulk states, the drumhead surface states appear a little fuzzy. Different from the usual DNL semimetal, the drumhead surface states of YH$_3$ are independent of its surface orientations, i.e., the [001]- and [010]-orientated surface states are almost the same [19], because of its near spherical band crossing.

The collective plasmon excitation is calculated using the Lindhard function and random phase approximation [19]. The electron energy loss spectrum (EELS), whose broad peaks indicate the plasmons [28], is shown in Figs. 2(g) and (h) using a $T = 300$ K Fermi-Dirac distribution. Due to the isotropy, we only show the wave vector along the $x$ direction. At the long-wavelength limit, the plasmon frequency almost keeps a constant value independent of $q$ and $n$, in good agreement with Eq. (2c). Such exotic plasmon excitations reflect the peculiarity of PDNS; the unusual stable plasmon exciation does not depend on the carrier concentrations. The constant excitation energy of YH$_3$ is 28 meV, corresponding to THz spectroscopy. It is noted that the $q$- and $n$-independent plasmon frequency is only applicable to the energy range of linear dispersion [19]. As indicated in Eq. (2c), the constant plasmon frequency is predicted to be determined by $v_F$ and $k_0$ ($\omega \propto \sqrt{v_F k_0^2}$), which is again confirmed by our extensive calculations including different external strains [19]. The constancy and tunability of plasmon frequency in the PDNS semimetal imply interesting applications in THz plasmonics.

As a candidate material for type-II PDNS semimetal, α-Si$_3$N$_2$ [29] adopts the cubic structure having the space group $Pm\bar{3}m$ (No. 221), as shown in Fig. 3(a). α-Si$_3$N$_2$ is a normal semiconductor [19], whose VBM and CBM belong to the $A_{2g}$ and $A_{1g}$ representations of $O_h$ point group, respectively. Based on our analysis (Table I), a type-II PDNS phase can be achieved when these two bands are crossed. We found that a sufficiently large triaxial compressive strain of $\varepsilon < -5\%$ (corresponds to a hydrostatic pressure of ~30 GPa) can induce this desired phase [19], as shown in Fig. 3(b). Near the Fermi level, band crossing persists along any arbitrary direction around Γ [Fig. 3(c)]. Importantly, although the two crossing bands have the same parities, the eigenvalues of six mirror planes for these two bands are of opposite sign [labelled in Fig. 3(c)]. Thus, α-Si$_3$N$_2$ is a type-II PDNS with multiple DNLs protected by the mirror symmetries (the calculated $\Delta c=1$). As expected, it has a constant DOS in the energy range of nearly linear dispersion [Fig. 3(c)], a spherical Fermi surface [Fig. 3(d)], surface-independent drumhead surface states [Figs. 3(e) and (f)], and $q$- and $n$-independent constant plasmon frequency [Figs. 3(g) and (h)]. The PDNS phase in α-Si$_3$N$_2$ is robust with a very small SOC gap (< 0.1 meV) [19].

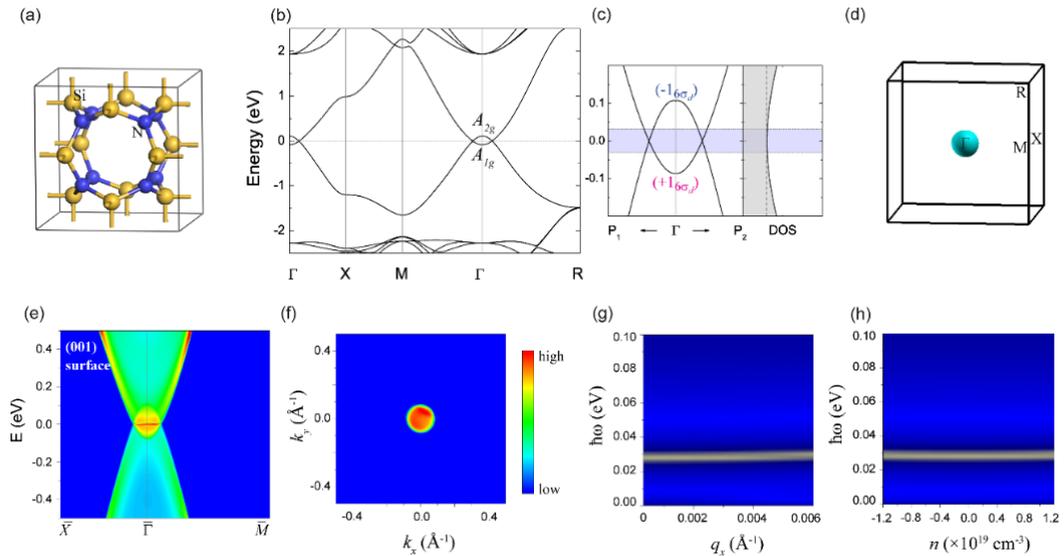

Figure 3. (a)-(h) Same as Figure 2 but for type-II α-Si$_3$N$_2$.

As a high-symmetry-required state, the PDNS semimetal can be considered as the "parent phase" for other gapped and gapless topological states. For instance, certain perturbations may tune the HIT transforming a PDNS semimetal into a nodal-line semimetal [19]; a sufficiently large SOC may convert a PDNS semimetal into a DNP semimetal or a topological insulator [19]. Moreover, because of the finite DOS in the linear band crossing region [Fig. 1(c)], Coulomb repulsion might drive the PDNS phase to induce various quantum orders [30]. Especially, the existence of superconductivity in $M$H$_3$ under pressure (strain) [24,25] may provide a unique platform to study the interplay between the PDNS fermions and superconductivity.

**Acknowledgements:** The authors thank L. Zhang, L. Lim, S.W. Gao and S.-H. Wei for helpful discussions. J.W. and B.H. acknowledge the support from NSFC (Grant No. 11574024) and NSAF U1530401. K.J. and F.L. acknowledge the support from US-DOE (Grant No. DE-FG02-04ER46148). Y.L., X.S. and W.D. acknowledge support from MOST of China (Grant No. 2016YFA0301001) and NSFC (Grants No. 11674188 and No. 11334006). Part of the calculations were performed at Tianhe2-JK at CSRC.